\documentclass[preprint,preprintnumbers,nofootinbib]{revtex4-1}
\usepackage{graphicx}% Include figure files
\usepackage{dcolumn}% Align table columns on decimal point
\usepackage{bm}% bold math
\usepackage{amsmath}
\usepackage{amsfonts} 
\usepackage{latexsym}
\usepackage{bbm}
\usepackage{color}
\usepackage{amssymb}
\usepackage{amsthm}
\usepackage{epsfig}
\usepackage{hyperref}
\hypersetup{%
setpagesize=false,
 bookmarksnumbered=true,%
 bookmarksopen=true,%
 colorlinks=true,%
 linkcolor=blue,
 citecolor=red,
}

\begin{document}

\preprint{\textbf{OCHA-PP-366}}

\title{Electroweak phase transition in a complex singlet extension of the Standard Model with degenerate scalars 
}

\author{Gi-Chol Cho$^1$}
\email{cho.gichol@ocha.ac.jp}
\author{Chikako Idegawa$^2$}
\email{c.idegawa@hep.phys.ocha.ac.jp}
\author{Eibun Senaha$^{3,4}$}
\email{eibunsenaha@vlu.edu.vn}
\affiliation{$^1$Department of Physics, Ochanomizu University, Tokyo 112-8610, Japan}
\affiliation{$^2$Graduate school of Humanities and Sciences, Ochanomizu University, Tokyo 112-8610, Japan}
\affiliation{$^3$Subatomic Physics Research Group, Science and Technology Advanced Institute, Van Lang University, Ho Chi Minh City, Vietnam}
\affiliation{$^4$Faculty of Technology, Van Lang University, Ho Chi Minh City, Vietnam}
\bigskip

\date{\today}

\begin{abstract}
We study the feasibility of strong first-order electroweak phase transition (EWPT) in a degenerate-scalar scenario of a complex singlet extension of the Standard Model, in which a mass of an additional scalar is nearly degenerate with that of the Higgs boson, 125 GeV. 
This scenario is known to provide an exquisite solution for circumventing constraints from dark matter direct detection experiments due to cancellations between two scattering amplitudes mediated by two scalars. In the analysis of EWPT, we employ two gauge-invariant calculation schemes on the scalar potential and two familiar resummation methods in evaluating one-loop (gauge dependent) effective potential. 
We point out that one of the conditions for the strong first-order EWPT is incompatible with the known suppression mechanism of a dark matter cross-section scattering off the nucleons. Nevertheless, we find that strong first-order EWPT is still possible in the degenerate-scalar scenario by dodging dark matter constraints differently.

\end{abstract}

\maketitle

%%%%%%%%%%%%%%%%%%%%%%%%%%%%%%%%%%%%%%%%%%%%%%%%%%%%%
% Introduction
%%%%%%%%%%%%%%%%%%%%%%%%%%%%%%%%%%%%%%%%%%%%%%%%%%%%%
\section{Introduction}

Clarifying the origin of baryon asymmetry of the Universe (BAU) is one of the long-standing problems in particle physics and cosmology. Cosmological observations from big-bang nucleosynthesis and cosmic microwave background consistently find the baryon-to-photon ratios, {\it i.e.}, $\eta^{\text{BBN}}=(5.8-6.5)\times10^{-10}$ and $\eta^{\text{CMB}}=(6.105\pm0.055)\times 10^{-10}$ at 95\% CL, respectively~\cite{Zyla:2020zbs}.
To generate BAU, one needs the so-called Sakharov's conditions~\cite{Sakharov:1967dj}: (i) baryon number violation, (ii) C and CP violation, and (iii) out of equilibrium. In principle, all the conditions can be satisfied in the Standard Model (SM) so that BAU can arise via electroweak baryogenesis mechanism~\cite{Kuzmin:1985mm} (for some reviews, see, e.g., Refs.~\cite{Quiros:1994dr,*Rubakov:1996vz,*Funakubo:1996dw,*Riotto:1998bt,*Trodden:1998ym,*Bernreuther:2002uj,*Cline:2006ts,*Morrissey:2012db,*Konstandin:2013caa,*Senaha:2020mop}). 
As widely known, however, the CP-violating effect coming from Cabibbo-Kobayashi-Maskawa matrix is too tiny to fuel BAU~\cite{Gavela:1993ts,*Gavela:1994dt,*Huet:1994jb,*Konstandin:2003dx,*Senaha:2020mop}, and out of equilibrium process cannot occur due to lack of first-order electroweak phase transition (EWPT) with the observed 125 GeV Higgs boson~\cite{Kajantie:1996mn,*Rummukainen:1998as,*Csikor:1998eu,*Aoki:1999fi}, calling for new physics. 
The necessity of new physics is also indicated because SM does not have a proper dark matter (DM) candidate. Though a plethora of DM candidates is conceivable, much attention has been drawn to weakly interacting massive particles which can naturally explain the observed DM relic density $\Omega_{\text{DM}}h^2 =0.1200\pm 0.001$~\cite{Zyla:2020zbs}. Currently, the statistical property of DM remains unknown, and both fermionic and bosonic DM cases are being actively investigated.
Among various new physics models, extensions of SM with multi-Higgs fields have been studied extensively as the simplest framework that can resolve two cosmological problems mentioned above.  
For example, the SM with a complex singlet scalar (CxSM)~\cite{Barger:2008jx,Barger:2010yn,Gonderinger:2012rd} can offer both the strong first-order EWPT and DM candidate simultaneously (for devoted studies, see, {\it e.g.}, Refs.~\cite{Basler:2020nrq,Jiang:2015cwa,Chiang:2017nmu,Chen:2019ebq}).
Despite sedulous experimental efforts by LHC, XENON1T~\cite{Aprile:2017iyp,*Aprile:2018dbl,*Aprile:2020vtw} etc., a hint of new physics has not emerged yet, which places severe constraints on parameter space consistent with strong first-order EWPT as well as DM mass and couplings to the SM sector. A solution to evade the DM direct detection constraints requires an additional singlet scalar nearly degenerating with the SM-like Higgs boson mass. In this case, a spin-independent (SI) cross section of DM with the nucleons ($\sigma_{\text{SI}}$) could be suppressed by a built-in cancellation mechanism regardless of Higgs-DM couplings~\cite{Kim:2008pp, Baek:2011aa,Kainulainen:2015sva,Azevedo:2018oxv,Azevedo:2018exj,Ishiwata:2018sdi,Huitu:2018gbc,Alanne:2018zjm,Cline:2019okt}. A recent detailed study in the CxSM~\cite{Abe:2021nih} revealed that the singlet scalar with a nearly 125 GeV mass (called \textit{degenerate-scalar scenario}) is yet consistent with all the experimental data, and that parameter space accommodating the correct DM relic density is wide open. 
On the other hand, although there are many studies on EWPT in the CxSM, such a degenerate-scalar scenario and possible parameter space have not been properly searched so far. 
In this paper,  we clarify the feasibility of strong first-order EWPT in the CxSM, focusing exclusively on the degenerate-scalar scenario. We survey EWPT using two gauge-invariant schemes: a tree-level potential with thermal masses (HT scheme) and Patal-Ramsey-Musolf (PRM) scheme~\cite{Patel:2011th}, as well as ordinary gauge-variant one-loop effective potentials with Parwani~\cite{Parwani:1991gq} or Carrington-Arnold-Espinosa (CAE) resummation scheme~\cite{Carrington:1991hz, Arnold:1992rz}. 
We note that the differences between these four methods for EWPT in CxSM have not been compared before.
The paper is organized as follows. In Sec.~\ref{sec:model}, we introduce the CxSM and describe the degenerate-scalar scenario in some detail. In Sec.~\ref{sec:ewpt}, we discuss the possibility of first-order EWPT in the degenerate-scalar scenario and clarify characteristics of the parameter space. Numerical results are presented in Sec.~\ref{sec:results}, and Sec.~\ref{sec:con_dis} is devoted to conclusion and discussions.

%%%%%%%%%%%%%%%%%%%%%%%%%%%%%%%%%%%%%%%%%%%%%%%%%%%%%
%Model
%%%%%%%%%%%%%%%%%%%%%%%%%%%%%%%%%%%%%%%%%%%%%%%%%%%%%

\section{Model}\label{sec:model}
The CxSM is the extension of the SM by adding the complex SU(2) gauge singlet scalar field. It is shown in Ref.~\cite{Barger:2008jx} that models with broken global U(1) symmetry can provide a pseudoscalar DM. 
Though various U(1) breaking terms are present in the potential, not all of them must address the strong first-order EWPT and viable DM. In this work, we consider a minimal and renormalizable model described by
\begin{align}
V_0 = \frac{m^2}{2}H^\dagger H+\frac{\lambda}{4}(H^\dagger H)^2
	+\frac{\delta_2}{2}H^\dagger H|S|^2+\frac{b_2}{2}|S|^2+\frac{d_2}{4}|S|^4
	+\bigg(a_1S+\frac{b_1}{4}S^2+{\rm H.c.}\bigg),
\end{align}
where both $a_1$ and $b_1$ break the global U(1) symmetry and the former is needed to avoid an unwanted $Z_2$ symmetry that could cause a domain wall problem once it is spontaneously broken. It should be noted that the U(1) breaking parameters must not generate a complex phase that induces a mixing between scalar and pseudoscalar components of $S$ in order to maintain the stability of DM.
The two scalar fields are parametrized as
\begin{align}
H =
	\left(
		\begin{array}{c}
		G^+ \\
		\frac{1}{\sqrt{2}}(v+h+iG^0)
		\end{array}
	\right),\quad
S =\frac{1}{\sqrt{2}}(v_S+s+i\chi),
\end{align}
where $v~(\simeq 246~\text{GeV})$ and $v_S$ are vacuum expectation values (VEVs), $h$ is the Higgs field that can mix with a singlet scalar $s$. $G^0$ and $G^\pm$ are Nambu-Goldstone fields, and a pseudoscalar $\chi$ is the DM candidate. 
First derivatives of $V_0$ with respect to $h$ and $s$ are respectively given by
\begin{align}
	\frac{1}{v}\left\langle\frac{\partial V_0}{\partial h}\right\rangle 
&= \frac{m^2}{2}+\frac{\lambda}{4}v^2+\frac{\delta_2}{4}v^2_{S}=0, \label{tad_h} \\
	\frac{1}{v_{S}}\left\langle\frac{\partial V_0}{\partial s}\right\rangle
&= \frac{b_2}{2}+\frac{\delta_2}{4}v^2
	+\frac{d_2}{4}v^2_{S}+\frac{\sqrt{2}a_1}{v_{S}}+\frac{b_1}{2} = 0, \label{tad_s}
\end{align}
where $\langle\cdots\rangle$ defines an operation such that fluctuation fields are taken to be zero.
Nonzero $v_S$ is enforced by $a_1\neq0$. For convention, we take $v_S>0$. With the above tadpole conditions,  the mass matrix in the basis $(h, s)$ is cast into the from  
\begin{align}
\mathcal{M}_S^2 &= 
\left(
\begin{array}{cc}
	\lambda v^2/2 & \delta_2vv_{S}/2 \\
	\delta_2vv_{S}/2 & \quad d_2v_{S}^2/2-\sqrt{2}a_1/v_{S}
\end{array}
\right),\label{Ms_tree}
\end{align}
which is diagonalized by an orthogonal matrix $O(\alpha)$ as
\begin{align}
O(\alpha)^T\mathcal{M}_S^2O(\alpha)  =
\left(
	\begin{array}{cc}
	m_{h_1}^2 & 0 \\
	0 & m_{h_2}^2
	\end{array}
\right), \quad
O(\alpha) = 
\left(
	\begin{array}{cc}
	\cos\alpha & -\sin\alpha \\
	\sin\alpha & \cos\alpha
	\end{array}
\right),
\end{align}
with $\alpha$ being the mixing angle such that $(h, s)^T=O(\alpha)(h_1, h_2)^T$.
The DM mass is calculated by the second derivative of $V_0$ with respect to $\chi$, which has the form
\begin{align}
m_\chi^2 = \frac{b_2}{2}-\frac{b_1}{2}\
	+\frac{\delta_2}{4}v^2+\frac{d_2}{4}v_{S}^2
	= -\frac{\sqrt{2}a_1}{v_{S}}-b_1,\label{mA_tree}
\end{align}
where the tadpole condition (\ref{tad_s}) is used in the second equality.  
We trade the original parameters $\{m^2$, $b_2$, $\lambda$, $d_2$, $\delta_2$, $b_1\}$ with $\{v$, $v_{S}$, $m_{h_1}$, $m_{h_2}$, $\alpha$, $m_\chi \}$ while retaining $a_1$ as an input. We set  $m_{h_1}=125$ GeV hereafter. In this model, $h_1$ couplings to the gauge bosons and fermions are scaled by $\cos\alpha$ while $h_2$ couplings to those particles by $-\sin\alpha$. Clearly, $\alpha\to 0$ corresponds to the SM-like limit which is compatible with current experimental data. As discussed below, however, this is not only the case that can mimic the SM.
Recent DM direct detection experiment by XENON1T puts an upper bound on the spin-independent (SI) cross section of the DM scattering off nucleons ($\sigma_{\text{SI}}$)~\cite{Aprile:2017iyp,*Aprile:2018dbl,*Aprile:2020vtw}, thereby constraining aforementioned model parameters. In the CxSM, both $h_1$ and $h_2$ get involved in the leading-order $t$-channel scattering process, and $\sigma_{\text{SI}}$ in the limit of vanishing momentum transfer has the following proportionality 
\begin{align}
\sigma_{\text{SI}}\propto \sin^2\alpha\cos^2\alpha \left(\frac{1}{m_{h_1}^2}-\frac{1}{m_{h_2}^2}\right)^2\frac{a_1^2}{v_S^4}.
\label{SIsigma}
\end{align}
One can see that the cross section can be highly suppressed for $m_{h_1}\simeq m_{h_2}$, irrespective of the other parameters.\footnote{It is shown that $\sigma_{\text{SI}}\propto m_{h_1}^2-m_{h_2}^2$ even at one-loop level (see., {\it e.g.}, Ref.~\cite{Azevedo:2018exj}).}
 The relative minus sign between the $h_1$ and $h_2$ contributions is attributed to the orthogonality of the mixing matrix $O(\alpha)$ and independent of the sign convention of $\alpha$. Note that the allowed ranges of $\alpha$ and $m_{h_2}$ are restricted considerably by Higgs coupling measurements and direct searches of extra scalars at LHC. As emphasized in Ref.~\cite{Abe:2021nih}, however, the parameter space where $|m_{h_1}-m_{h_2}|\lesssim3$ GeV is less constrained by the current LHC data~\cite{Khachatryan:2014ira}, allowing even the maximal mixing angle $|\alpha|=\pi/4$. 
Therefore, such a degenerate scalar scenario can evade the DM direct detection constraints.
It should be noted that although the small $\alpha$ could be a choice to circumvent the XENON1T bound, such a small mixing region would not be the right direction to go from the strong first-order EWPT point of view, as detailed in Sec.~\ref{sec:ewpt}.
We notice in passing that, as first pointed out in Ref.~\cite{Barger:2010yn} and emphasized in Ref.~\cite{Gross:2017dan}, $\sigma_{\text{SI}}$ could also vanish in the limit of $a_1=0$, which is another way to dodge the DM direct detection constraints. However, one should somehow avoid the domain wall problem mentioned earlier. It is known that the strong first-order EWPT is feasible for $a_1=0$ (see, {\it e.g.}, Ref.~\cite{Chiang:2017nmu}).
%%%%%%%%%%%%%%%%%%%%%%%%%%%%%%%%%%%%%%%%%%%%%%%%%%%%%
%		            Electroweak phase transition in the degenerate scalar-scenario
%%%%%%%%%%%%%%%%%%%%%%%%%%%%%%%%%%%%%%%%%%%%%%%%%%%%%
\section{Electroweak phase transition in the degenerate-scalar scenario}\label{sec:ewpt}
Now we discuss EWPT in the degenerate-scalar scenario. 
We take the Landau gauge $\xi=0$ with $\xi$ representing a gauge-fixing parameter. 
Denoting the classical background fields of the Higgs doublet and singlet as $\langle H\rangle =(0~~\varphi)^T/\sqrt{2}$ and $\langle S\rangle=\varphi_S/\sqrt{2}$, the effective potential at one-loop level takes the form
\begin{align}
V_1(\varphi, \varphi_S; T) = \sum_{i}n_i
\left[V_{\rm CW}(\overline{m}_i^2)
	+\frac{T^4}{2\pi^2}I_{B,F}\left(\frac{\overline{m}_i^2}{T^2}\right)
\right],\label{V1}
\end{align}
where $i=h_{1,2}, \chi, W, Z$ and $t$, and degrees of freedom and statistics of each particle is respectively given by $n_{h_1}=n_{h_2}=n_\chi = 1$, $n_W=6$, $n_Z = 3$ and $n_t=-12$.\footnote{We do not take account of the Nambu-Goldstone boson contributions which should be taken with special care to avoid infrared divergences when renormalizing $V_{\text{CW}}$ adopted here. Our conclusion is not affected by this omission.} $V_{\rm CW}$ is the zero temperature piece while $I_{B,F}$ are the nonzero temperature counterpart, which are respectively given by
\begin{align}
V_{\text{CW}}(\overline{m}_i^2) &= \frac{\overline{m}_i^4}{64\pi^2}\left(\ln\frac{\overline{m}_i^2}{\overline{\mu}^2}-c_i\right), \\
I_{B,F}(a^2) &= \int_0^\infty dx~x^2\ln\Big(1\mp e^{-\sqrt{x^2+a^2}}\Big),
\end{align}
where $\overline{\mu}$ denotes a renormalization scale and $c_i=3/2$ for scalars and fermions while $c_i=5/6$ for gauge bosons. In our loop calculations except for the PRM scheme detailed below\footnote{The detail of the PRM scheme is given after investigating EWPT analytically by use of a high-temperature potential.}, 
$V_{\text{CW}}$ is renormalized in such a way that the tree-level relations of the tadpole conditions and two scalar masses and its mixing given in Eqs.~(\ref{tad_h})-(\ref{Ms_tree}) hold even at one-loop level. { In this case, $\bar{\mu}$ is fixed to some value.} $\overline{m}_i$ denote the field-dependent masses of the species $i$ and their values in the vacuum represent $m_i$. $I_B$ with the upper sign is the thermal function for bosons while $I_F$ with the lower one is that for fermions. At high temperature, where $\overline{m}^2\ll T^2$, those thermal functions can be expanded in powers of $a^2$, which makes it easy to see how first-order EWPT is realized and which parameters play a major role for that.
In the presence of the singlet field, the potential barrier can be induced by the doublet-singlet mixing that exists
at the tree level. For illustration, we first examine EWPT using a high-temperature (HT) potential defined as
\begin{align}
V^{\text{HT}}(\varphi, \varphi_S;T) = V_0(\varphi, \varphi_S)+\frac{1}{2}\left(\Sigma_H\varphi^2+\Sigma_S\varphi_S^2\right)T^2,
\label{VHT}
\end{align}
where
\begin{align}
\Sigma_H = \frac{\lambda}{8}+\frac{\delta_2}{24}+\frac{3g_2^2+g_1^2}{16}+\frac{y_t^2}{4}, \quad 
\Sigma_S = \frac{\delta_2+d_2}{12},
\end{align}
with $g_2$, $g_1$ and $y_t$ being $\text{SU(2)}_L$, $\text{U(1)}_Y$ and top Yukawa couplings, respectively.
Note that $V^{\text{HT}}$ is not only simple but free from gauge dependence thanks to the gauge-invariant thermal masses.\footnote{In this study, we classify the calculation using $V^{\text{HT}}$ into gauge-invariant scheme category though it is merely the high-temperature expanded one-loop effective potential (\ref{V1}) to leading order.} With this potential, it is possible to derive the analytic expressions of the critical temperature $T_C$ at which the potential has two degenerate minima and VEV of the Higgs doublet at $T_C$, which is denoted by $v_C$. 
For electroweak baryogenesis to work, a baryon number-violating process must be suppressed after EWPT, which is realized by first-order phase transition associated with bubble nucleations and expansions. As long as supercooling is moderate, $T_C$ defined above would not be much different from a nucleation temperature, and hence decoupling of the baryon number-violating process is required as
\begin{align}
\frac{v_C}{T_C}\gtrsim 1,\label{sph_dec}
\end{align}
where the lower bound is governed mostly by the energy of sphaleron configuration (for a recent study, see Ref.~\cite{Chiang:2017nmu}).
To make our analysis simpler, we first parametrize the two scalar fields using radial coordinates as
\begin{align}
\varphi = z\cos\gamma,\quad \varphi_S=z\sin\gamma+v_S^{\text{sym}},
\end{align}
where $v_S^{\text{sym}}$ denotes the minimum on the $\varphi_S$ axis, which is always nonzero due to $a_1\neq0$ in our investigation. In the case of first-order EWPT, $T_C$ and $v_C$ take the form
\begin{align}
T_C^2 & = \frac{1}{2(\Sigma_H+\Sigma_S t_{\gamma_C^{}}^2)} 
\bigg[
	-m^2-\frac{(v_{SC}^{\text{sym}})^2\delta_2}{2} \nonumber\\
&\hspace{4cm}
-\left\{
	b_1+b_2+
\left(\frac{3d_2}{2}-\frac{(\delta_2+d_2t_{\gamma_C^{}}^2)^2}{\lambda+2\delta_2t_{\gamma_C^{}}^2+d_2t_{\gamma_C^{}}^4}\right)(v_{SC}^{\text{sym}})^2	
\right\}t_{\gamma_C^{}}^2
\bigg], \label{Tcfull} \\
v_C & = \frac{-2t_{\gamma_C^{}}{v_{SC}^{\text{sym}}}
	(\delta_2+d_2t_{\gamma_C^{}}^2)}
	{\lambda +2\delta_2t_{\gamma_C^{}}^2 +d_2t_{\gamma_C^{}}^4}, \label{vcfull}
\end{align}
with 
\begin{align}
t_{\gamma_C^{}} = \frac{\sin\gamma(T_C)}{\cos\gamma(T_C) }= \frac{v_{SC}-v_{SC}^{\text{sym}}}{v_C}, \quad
v_C = \lim_{T\nearrow T_C}v(T), \quad 
v_{SC}=\lim_{T\nearrow T_C}v_S(T), \quad 
v_{SC}^{\text{sym}}=\lim_{T\searrow T_C}v_S(T), 
\end{align}
where $T\nearrow T_C$ and $T\searrow T_C$ are defined such that $T$ approaches $T_C$ from below and above, respectively. 
For $t_{\gamma_C^{}}\ll 1,$\footnote{Though our numerical results presented in Sec.~\ref{sec:results} satisfy $t_{\gamma_C^{}}<1$, its magnitude is not far smaller than unity. Therefore, the following argument in this section would be valid qualitatively and quantitative estimates need the full expressions~Eqs.~(\ref{Tcfull}) and (\ref{vcfull}).} $T_C$ and $v_C$ can be approximated as~\cite{Chiang:2017nmu}
\begin{align}
T_C &\simeq \sqrt{\frac{1}{2\Sigma_H}\left(-m^2-\frac{(v_{SC}^{\text{sym}})^2}{2}\delta_2\right)}, 
\label{Tcapp}
\\
v_C&\simeq \sqrt{\frac{2\delta_2(v_{SC}^{\text{sym}})^2}{\lambda}
	\left(1-\frac{v_{SC}}{v_{SC}^{\text{sym}}}\right)}. 
\label{vcapp}
\end{align}
As seen from Eq.~(\ref{Tcapp}), $T_C$ would get lowered if $\delta_2$ is positive and sizable. 
By expressing the Lagrangian parameters in terms of the input parameters, one finds
\begin{align}
\delta_2 &= \frac{2}{vv_{S}}(m_{h_1}^2-m_{h_2}^2)\sin\alpha\cos\alpha. 
\label{del2}
\end{align}
To have large $\delta_2$ in the degenerate scenario ($m_{h_1}\approx m_{h_2}$), 
$v_S$ has to be small in order to compensate the small mass difference between $h_1$ and $h_2$. Moreover, the maximal mixing $|\alpha|=\pi/4$ is also preferred. 
Note that $\delta_2$ is invariant under the transformation $m_{h_1}^2-m_{h_2}^2\to -(m_{h_1}^2-m_{h_2}^2)$ and $\alpha\to -\alpha$, implying that $\delta_2$ can always be made positive.
For sizable $\delta_2$, 
$v_C$ in Eq.~(\ref{vcapp}) would also be enhanced with an amplification factor $(v_{SC}^{\text{sym}})^2(1-v_{SC}/v_{SC}^{\text{sym}})$. Note that $v_{SC}^{\text{sym}}$ is determined by the cubic equation 
\begin{align}
(v_{SC}^{\text{sym}})^3+Av_{SC}^{\text{sym}}+B = 0,\label{cubic_vStil}
\end{align}
with $A = 2(b_1+b_2+2\Sigma_S)/d_2$ and $B = 4\sqrt{2}a_1/d_2$. One can find that real solutions, if they exist, are scaled by $1/\sqrt{d_2}$, which indicates that smaller $d_2$ would be supportive of larger $v_C$. 
When $v_S$ is small, which is prefered for sizable $\delta_2$, 
it is easy to see that $d_2$ could get enhanced for $m_{h_1}^2 \approx m_{h_2}^2$ as 
\begin{align}
d_2 &= \frac{2}{v_{S}^2}
\left[
	m_{h_1}^2+(m_{h_2}^2-m_{h_1}^2)\cos^2\alpha+\frac{\sqrt{2}a_1}{v_{S}}
\right]
\simeq
\frac{2}{v_{S}^2}
\left[
	m_{h_1}^2+\frac{\sqrt{2}a_1}{v_{S}}
\right].
\end{align}
The moderate small value of $d_2$ discussed above determines the magnitude of $a_1$ and its sign, {\it i.e.}, $a_1<0$. Moreover, one can see that the sign of $m_{h_2}^2-m_{h_1}^2$ cannot be compensate by that of $\alpha$, which is in contrast with $\delta_2$. In our numerical analysis, we consider both $m_{h_1}<m_{h_2}$ and $m_{h_1}>m_{h_2}$ cases. 
Recapitulating the above argument, to have the strong first-EWPT in the degenerate scalar scenario, one needs 
\begin{itemize}
\item[(i)] large $\delta_2$ with a positive sign, {\it i.e.}, $|\alpha|\simeq \pi/4$ and $v_S<1$ GeV,
\item[(ii)] small $d_2$, {\it i.e.}, $a_1<0$ with its moderate absolute value.
\end{itemize}
 However, the sizes of $\delta_2$ and $d_2$ could be constrained by a global minimum condition at $T=0$. The energy difference between the electroweak vacuum prescribed by $(v, v_S)$ and the local vacuum on the $\varphi_S$ axis specified by $(0,v_S^{\text{sym}})$ is 
\begin{align}
\Delta E&=V_0(0,v_S^{\text{sym}})-V_0(v,v_S)\nonumber\\
& = \sqrt{2}a_1(v_S^{\text{sym}}-v_S)+\frac{1}{4}(b_1+b_2)\left((v_S^{\text{sym}})^2-v_S^2\right)
 	+\frac{d_2}{16}\left((v_S^{\text{sym}})^4-v_S^4\right) \nonumber\\
&\quad-\frac{m^2}{4}v^2-\frac{\lambda}{16}v^4-\frac{\delta_2}{8}v^2v_S^2,
\end{align}
which has to be positive. However, $\Delta E$ could be negative for $\delta_2\gg1$ and $d_2\ll1$ while other parameters are kept fixed, thereby placing the upper and lower bound on them, respectively. Note that $v_S$ has no limitation to its smallness from the condition $\Delta E>0$.
Even though the HT potential is of much use to discuss EWPT qualitatively, validity of the high-temperature expansions of $I_{B,F}$ would not been guaranteed as temperature goes down. Furthermore, one-loop contributions at $T=0$ would not be negligible for quantitative study. In order to incorporate higher-order corrections in a gauge-invariant way, we employ the PRM scheme~\cite{Patel:2011th}. In this formalism, $T_C$ and the corresponding VEVs at $T_C$ are determined separately, and the higher-order corrections can be taken into account based on the Nielsen-Fukuda-Kugo (NFK) identity~\cite{Nielsen:1975fs,Fukuda:1975di}
\begin{align}
\frac{\partial V_{\text{eff}}(\varphi,\xi)}{\partial \xi}  = -C(\varphi,\xi)\frac{\partial V_{\text{eff}}(\varphi,\xi)}{\partial \varphi},
\end{align}
where $C(\varphi,\xi)$ denotes some functional that is calculable order by order in perturbation theory. One can obtain the NFK identity to given order by expanding each term in the both sides in powers of $\hbar$. In our work, $T_C$ is determined to $\mathcal{O}(\hbar)$ using the following degeneracy condition~\cite{Patel:2011th}
\begin{align}
V_0(0, v_{S,\text{tree}}^{\text{sym}})+V_1(0, v_{S,\text{tree}}^{\text{sym}}; T)=V_0(v_{\text{tree}}, v_{S,\text{tree}})+V_1(v_{\text{tree}}, v_{S,\text{tree}}; T),\label{Tc_PRM}
\end{align}
where $v_{\text{tree}}$, $v_{S,\text{tree}}$, and $v_{S,\text{tree}}^{\text{sym}}$ are the tree-level VEVs at $T=0$. In the PRM scheme, after $T_C$ is obtained by the above procedure, $v_C$, $v_{SC}$, and $v_{SC}^{\text{sym}}$ are determined by use of $V^{\text{HT}}$. { In contrast to the other one-loop calculation performed here, the PRM scheme has an explicit $\bar{\mu}$ dependence coming from $V_1$.} It is noticed in Refs.~\cite{Chiang:2017nmu,Chiang:2018gsn} that the $\overline{\mu}$ dependence is rather large in the $\mathcal{O}(\hbar)$ calculation, and $\mathcal{O}(\hbar^2)$ contributions are necessary for the quantitative study. In the current investigation, however, we confine ourselves to the $\mathcal{O}(\hbar)$ calculation as a first step toward more complete analysis, and set $\overline{\mu}=m_t=172.76$ GeV as a reference point.
We have so far discussed two methods (HT and PRM) to evaluate the effective potential in the gauge invariant manner. 
For comparisons, we also evaluate strength of EWPT using the one-loop potential (\ref{V1}) with daisy resummations. As commonly done in the literature, { we employ two thermal resummation methods. One is called Parwani scheme~\cite{Parwani:1991gq} and the other is Arnold-Espinosa (AE) scheme~\cite{Arnold:1992rz} which is equivalent to the earlier study by Carrington~\cite{Carrington:1991hz} (so the CAE scheme hereafter).}\footnote{A more careful treatment based on dimensional reduction can be found in Refs.~\cite{Croon:2020cgk,Gould:2021oba}.} For the former, we replace $\overline{m}^2_i$ appearing in $I_{B}$ and $I_F$ with thermally corrected field-depend masses denoted as {$\overline{M}_i^2$}. For the latter, on the other hand, we add
\begin{align}
V_{\text{daisy}}(\varphi, \varphi_S; T) 
&=\sum_{\stackrel{i=h_{1,2},\chi}{W_L,Z_L,\gamma_L}^{}}-n_i\frac{T}{12\pi}
	\Big[\big(\overline{M}_i^2\big)^{3/2}
	-\big(\overline{m}_i^2\big)^{3/2}\Big]
\end{align}
to the one-loop effective potential $V_0+V_1$, where $\overline{m}_i^2$ in $V_1$ remain intact. $W_L$, $Z_L$, and $\gamma_L^{}$ are the longitudinal parts of the gauge fields whose degrees of freedom are $n_{W_L}/2=n_{Z_L}=n_{\gamma_L^{}}=1$, respectively. In principle, the transverse parts of the gauge fields also receive non-perturbative thermal corrections. Since the first-order EWPT is predominantly induced by the tree-level potential in our case, such a correction has little effect on $v_C/T_C$. Nevertheless, as demonstrated in Ref.~\cite{Funakubo:2020wah}, the correction could be of potential importance for the sphaleron energy, which is beyond the scope of the current paper.
Before showing our numerical results, we discuss the compatibility between strong first-order EWPT and the DM direct detection constraints. 
Using Eq.~(\ref{del2}), one can rewrite the SI DM cross section (\ref{SIsigma}) as
\begin{align}
\sigma_{\text{SI}}\propto 
%\sin^2\alpha\cos^2\alpha \left(\frac{1}{m_{h_1}^2}-\frac{1}{m_{h_2}^2}\right)^2\frac{a_1^2}{v_S^4}=
\frac{\delta_2^2v^2}{4m_{h_1}^4m_{h_2}^4}\frac{a_1^2}{v_S^2}.
\end{align}
It must be emphasized here that the core of the cancellation mechanism in the degenerate-scalar scenario is the suppression of $\delta_2$ owing to $m_{h_1}\simeq m_{h_2}$ with {\it moderate values} of $v_S$. However, as far as the strong first-order EWPT is concerned, $v_S$ must be smaller than 1 GeV to render $\delta_2$ bigger, so the small $\delta_2$ as the solution for evading XENON1T constraint is the no-go zone. Therefore, the cancellation by the mass degeneracy would not work in our case, and we explicitly quantify this statement in the next section.

%%%%%%%%%%%%%%%%%%%%%%%%%%%%%%%%%%%%%%%%%%%%%%%%%%%%%
%Results
%%%%%%%%%%%%%%%%%%%%%%%%%%%%%%%%%%%%%%%%%%%%%%%%%%%%%
\section{Results}\label{sec:results}

%----------------------------------------------------------------------------------------------------------------------------------
\begin{table}[t]
\center
\begin{tabular}{|c|c|c|c|c|c|c|c|}
\hline
Inputs & $v$ [GeV] & $m_{h_1}$ [GeV] & $m_{h_2}$ [GeV] & $\alpha$ [rad] & $a_1$ [GeV$^3$] & $v_S$ [GeV] & $m_\chi$ [GeV]  \\ \hline
BP1 & 246.22 &125 & 124 & $\pi/4$ & $-6576.17$ & 0.6 & 62.5  \\ \hline
BP2 & 246.22 &125 & 126 & $-\pi/4$ & $-6682.25$ & 0.6 & 62.5  \\ \hline\hline
Outputs & $m^2$ [GeV$^2$] & $b_1$ [GeV$^2$] & $b_2$ [GeV$^2$] & $\lambda$ & $a_1$ [GeV$^3$] & $d_2$ & $\delta_2$  \\ \hline
BP1 & $-(124.5)^2$ &$(107.7)^2$ & $-(178.0)^2$ & 0.511 & $-6576.17$ & 1.77 & 1.69  \\ \hline
BP2 & $-(125.5)^2$ &$(108.8)^2$ & $-(178.4)^2$ & 0.520 & $-6682.25$ & 1.70 & 1.59  \\ \hline
\end{tabular}
\caption{The first two rows are the input parameters in BP1 and BP2. Apart from $a_1$, those are converted to the Lagrangian parameters given in the last two rows, where the tree-level relations are used. The BPs  satisfy vacuum stability and perturbativity.}
\label{tab:BP}
\end{table}
%----------------------------------------------------------------------------------------------------------------------------------
%

%
We begin by presenting the DM relic density $\Omega_\chi h^2$ and SI scattering cross section with the nucleons $\sigma_{\text{SI}}$ in parameter space that satisfies the criteria for the strong first-order EWPT discussed in the previous section. For illustrative purposes, we consider two benchmark points (BPs) listed in Table~\ref{tab:BP}. The chosen points do not upset vacuum stability and perturbativity, let alone LHC constraints.  
For the moment, $m_\chi$ is treated as the varying parameter. 
In our study, we use a public code \texttt{micrOMEGAs}~\cite{Belanger:2020gnr} to calculate $\Omega_\chi h^2$ and $\sigma_{\text{SI}}$. The value of $\Omega_\chi h^2$ should not exceed the observed value~\cite{Zyla:2020zbs}
\begin{align}
\Omega_{\text{DM}}h^2 = 0.1200\pm 0.0012.\label{Oh2obs}
\end{align}
On the other hand, $\sigma_{\text{SI}}$ is constrained by DM direct detection experiments such as XENON1T. 
In the case of $m_\chi=30$ GeV, for instance, the maximum value is $\sigma_{\text{SI}}\simeq 4.1\times 10^{-47}~\text{cm}^2$~\cite{Aprile:2017iyp,*Aprile:2018dbl,*Aprile:2020vtw},
under the assumption that $\Omega_\chi=\Omega_{\text{DM}}$. In cases that $\Omega_\chi<\Omega_{\text{DM}}$, we scale $\sigma_{\text{SI}}$ as
\begin{align}
\widetilde{\sigma}_{\text{SI}} = \left(\frac{\Omega_\chi}{\Omega_{\text{DM}}}\right)\sigma_{\text{SI}},
\end{align}
and impose that it should not surpass the XENON1T bound.
Since there are no significant numerical differences between BP1 and BP2, we only present the results in BP1 below.
 
\begin{figure}
\center
\includegraphics[width=7.8cm]{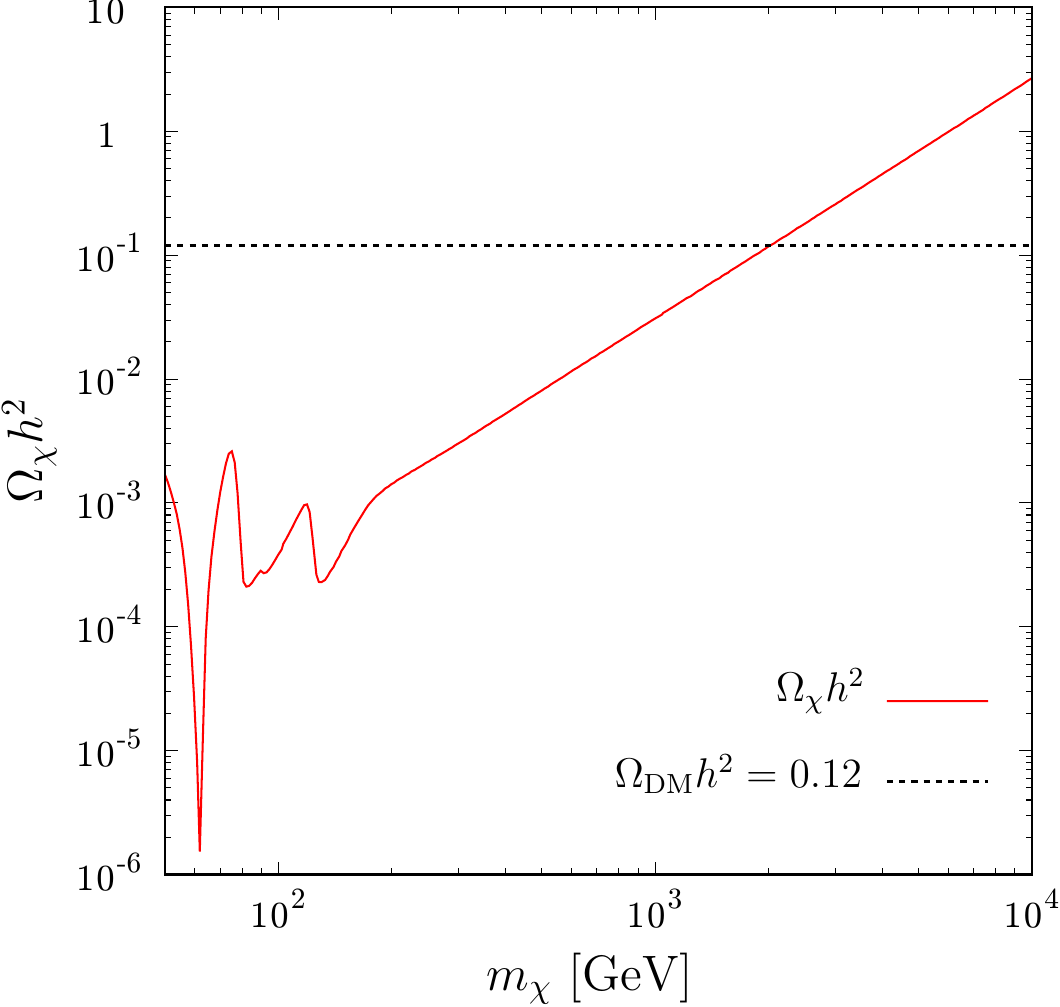}
\includegraphics[width=8cm]{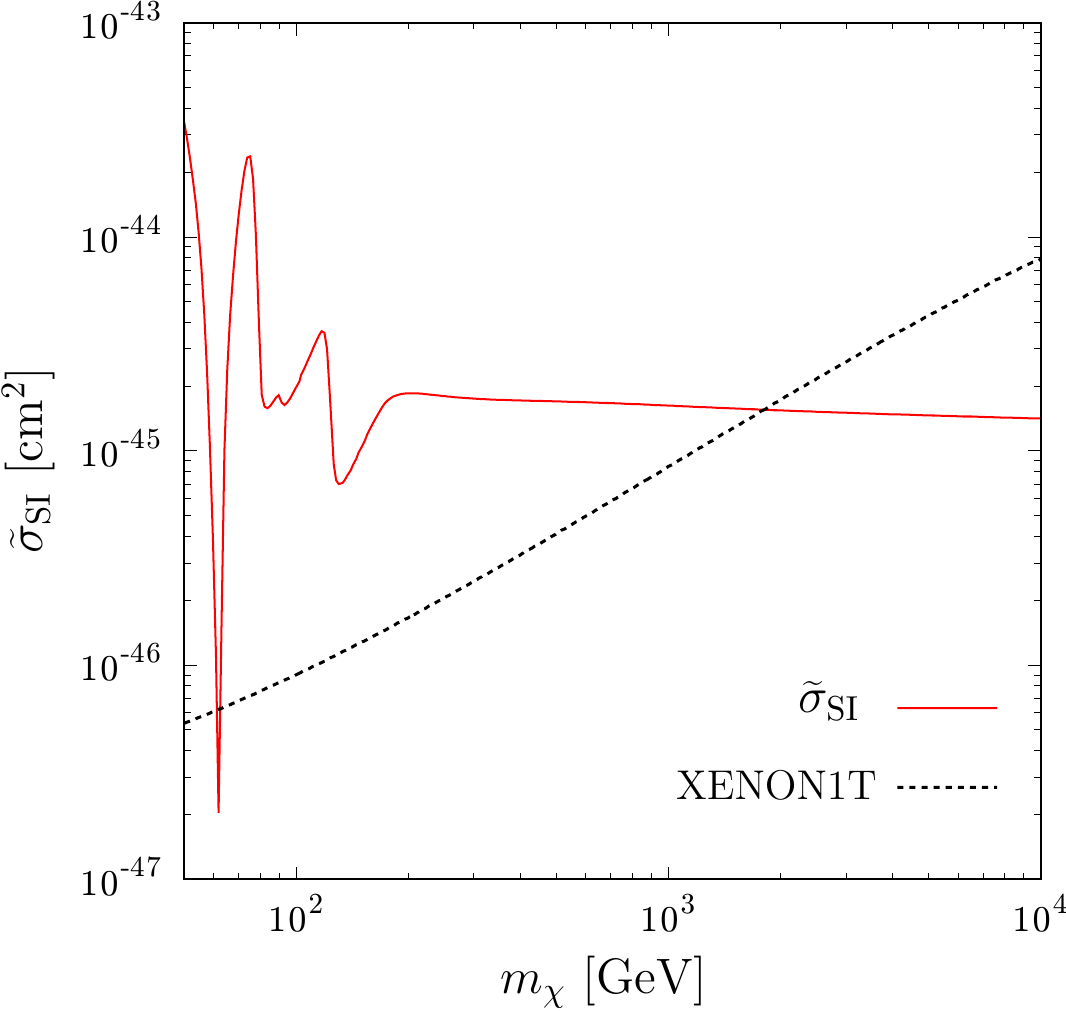}
\caption{The DM relic density $\Omega_\chi h^2$ (left panel) and scaled SI scattering cross section with the nucleons $\widetilde{\sigma}_{\text{SI}}$ (right panel) are plotted against the DM mass $m_\chi$. We take BP1 for other input parameters. The horizontal dotted lines in the left panel corresponds to the central value of the observed DM relic density, while the dotted curve in the right panel represents the XENON1T data.}
\label{fig:DM}
\end{figure}

$\Omega_\chi h^2$ is displayed as a function of $m_\chi$ in the left panel of Fig.~\ref{fig:DM}. The horizontal dotted line represents the central value of the observed DM relic density (\ref{Oh2obs}). We find that $\chi$ is not able to comprise of the entire DM abundance until its mass reaches around 2 TeV where $\Omega_\chi=\Omega_{\text{DM}}$. The sharp dip around $m_\chi=m_{h_1}/2~(=62.5~\text{GeV})$ reflects resonant enhancements of $s$-channel annihilation processes mediated by $h_1$ and $h_2$.

In the right panel of Fig.~\ref{fig:DM}, $\widetilde{\sigma}_{\text{SI}}$ against $m_\chi$ is shown, where the dotted curve indicates the XENON1T data. One can see that $\widetilde{\sigma}_{\text{SI}}$ around $m_\chi=62.5$ GeV is suppressed enough to avoid the XENON1T bound, which is due to the aforementioned dip of $\Omega_\chi h^2$. There exists another allowed region in which $m_\chi\gtrsim 2$ TeV. However, such a mass range would yield the overabundant DM, except for the marginal point $m_\chi\simeq 2020$ GeV in which $\Omega_\chi h^2=0.12$.

In summary, $\sigma_{\text{SI}}$ cannot be suppressed by the mass degeneracy of $h_1$ and $h_2$ since the corresponding $\delta_2$ remains sizable due to the smallness of $v_S$. Notwithstanding, the allowed regions still exist at around $m_\chi=62.5$ GeV and $2$ TeV. 
In what follows, we present a detailed analysis on EWPT taking $m_\chi=62.5$ GeV and comment on the heavier $\chi$ case at the end of this section.
% 

%
%---------------------------------------------------------------------------
%---------------------------------------------------------------------------
%

%
%---------------------------------------------------------------------------
\begin{figure}
\center
\includegraphics[width=8cm]{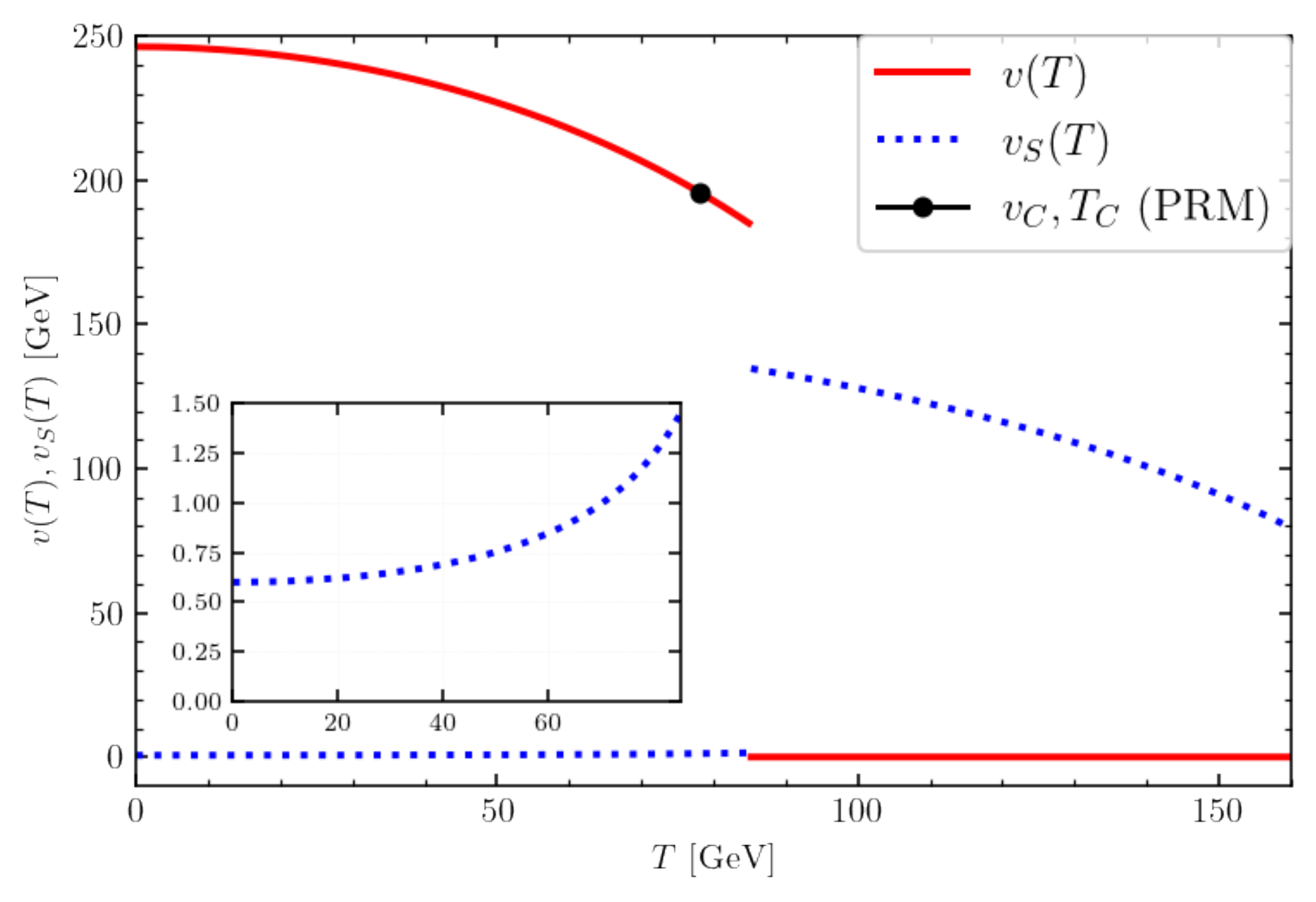}
\includegraphics[width=8cm]{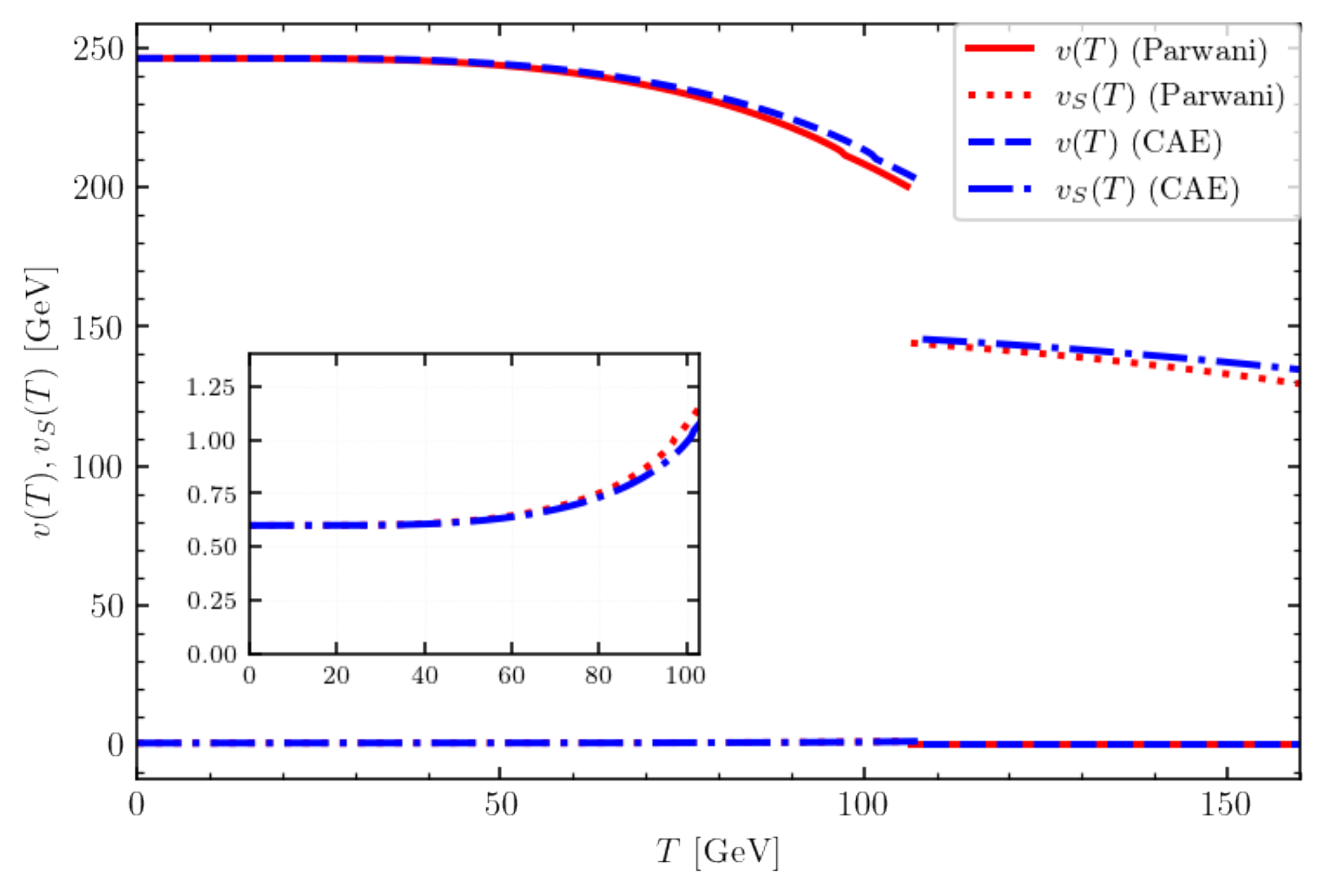}
\caption{$v(T)$ and $v_S(T)$ are shown as functions of the temperature $T$ in BP1. The HT potential is used in the left panel, while one-loop effective potentials with the Parwani and CAE resummation schemes are employed in the right panel. }
\label{fig:EWPT124}
\end{figure}
%---------------------------------------------------------------------------
%

%
%---------------------------------------------------------------------------
\begin{figure}
\center
\includegraphics[width=8cm]{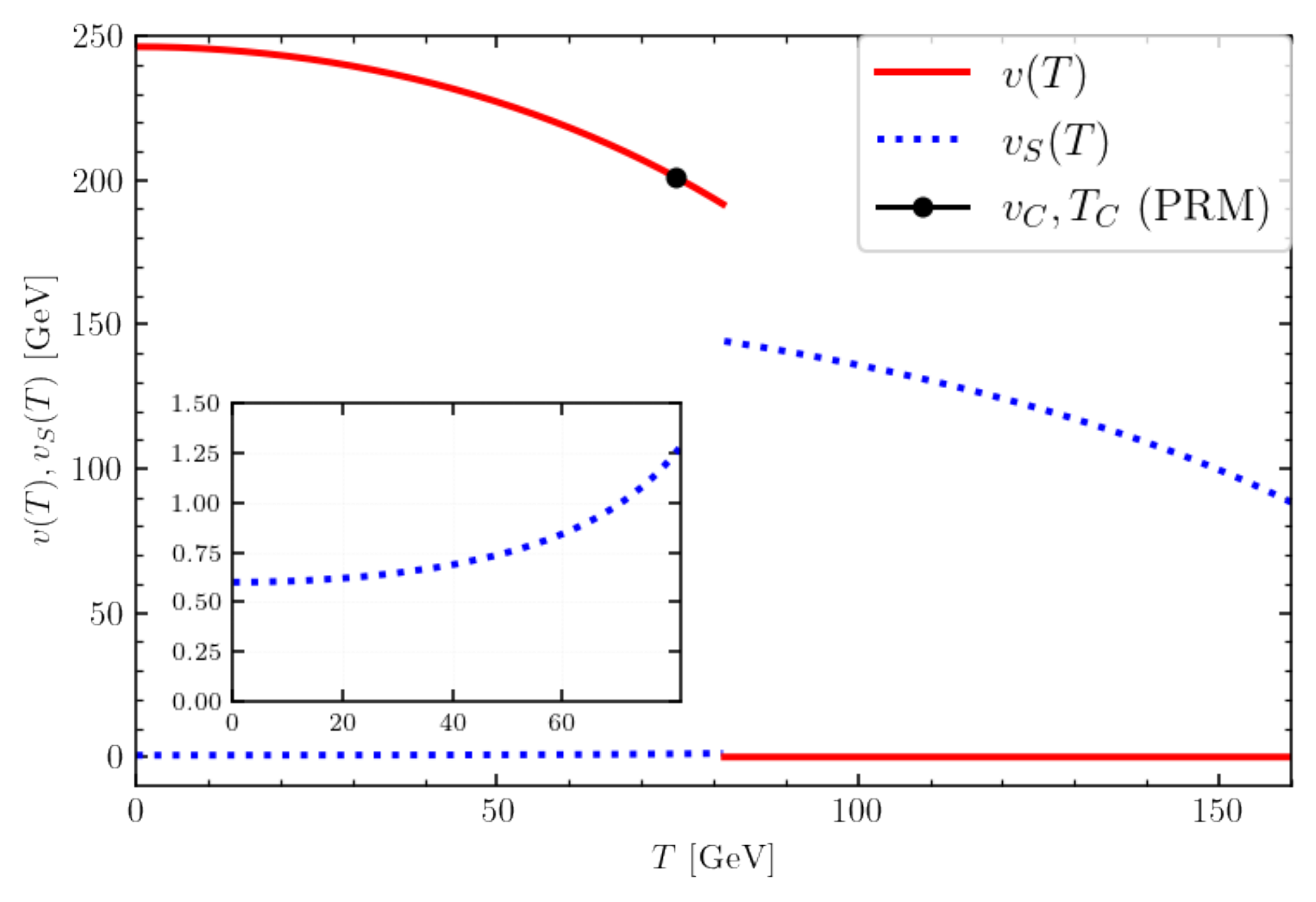}
\includegraphics[width=8cm]{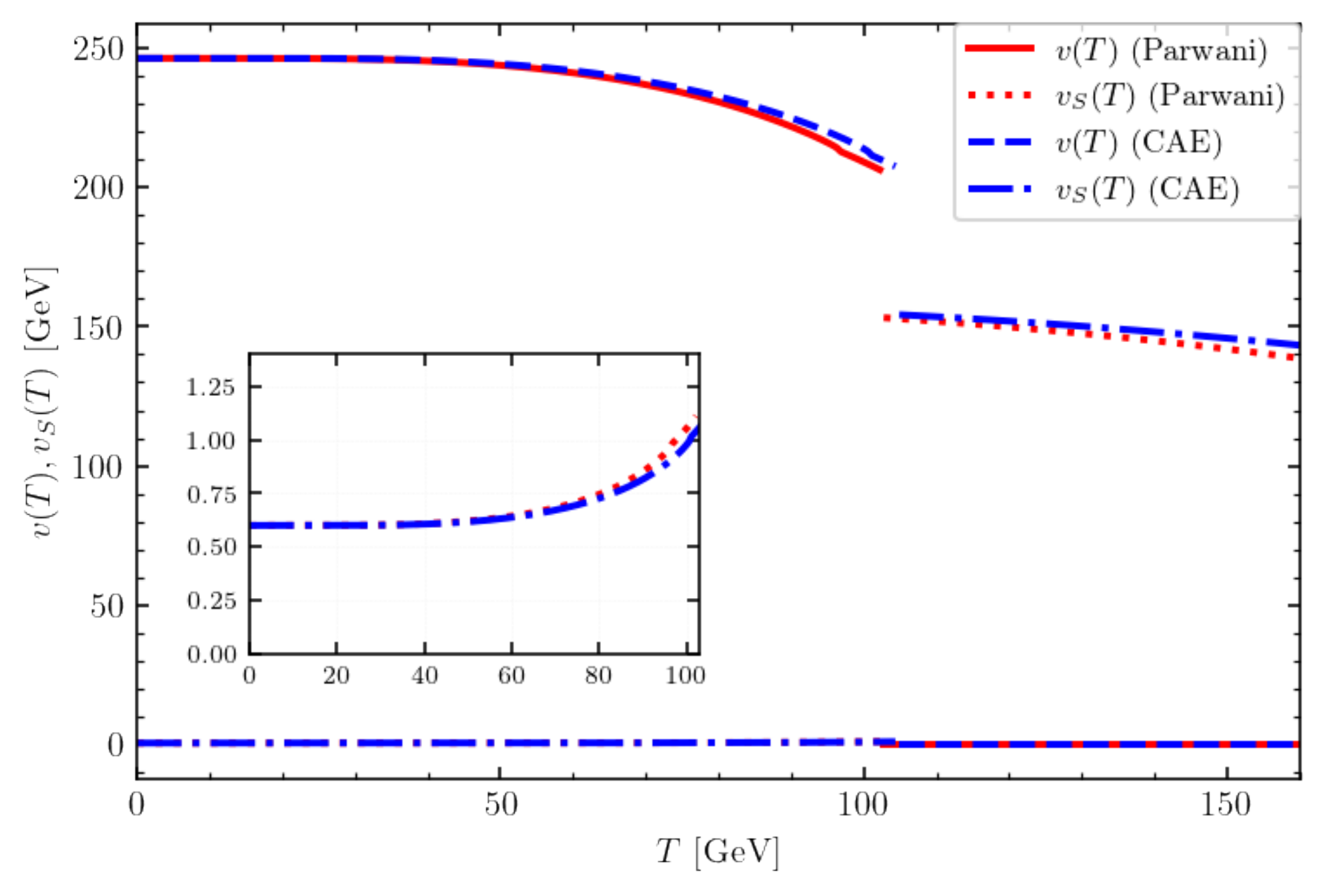}
\caption{$v(T)$ and $v_S(T)$ are shown as functions of the temperature $T$ in BP2. }
\label{fig:EWPT126}
\end{figure}
%---------------------------------------------------------------------------
%

%
In Fig.~\ref{fig:EWPT124}, we display the temperature dependences of VEVs in BP1. In the left panel, the HT potential is used, and $v(T)$ and $v_S(T)$ are shown by red solid and blue dotted curves, respectively, where magnified $v_S(T)$ is also plotted in the small window. As clearly seen, discontinuity exists in the temperature evolutions of VEVs, which determines $T_C^{\text{HT}}=85.3$ GeV, $v_C^{\text{HT}}=184.4$ GeV, $v_{SC}^{\text{HT}}=1.5$ GeV, and $v_{SC}^{\text{sym,HT}}=134.6$ GeV, and therefore the first-order EWPT is strong enough to satisfy the sphaleron decoupling condition $(\ref{sph_dec})$. As discussed in Sec.~\ref{sec:ewpt}, the first-order EWPT is accompanied by the sizable change of $v_S$, {\it i.e.,} $v_{SC}^{\text{sym}}\gg v_{SC}$, which is peculiar to the tree-potential driven first-order EWPT. 
In this plot, the results in the PRM scheme are also shown by the black solid circle, where $T_C^{\text{PRM}}=78.2$ GeV and $v_C^{\text{PRM}}=195.6$ GeV. 
Although it may not be able to make the definite statement until theoretical uncertainties are well under control, as noted in Sec.~\ref{sec:ewpt}, the obtained values in this scheme are also indicative of the strong first-order EWPT. 
In the right panel, one-loop effective potentials with the Parwani and CAE resummation schemes are employed to evaluate the temperature evolution of VEVs. The solid and dotted lines in red represent $v(T)$ and $v_S(T)$ in the Parwani scheme, while the dashed and dotted lines in blue denote ones in the CAE scheme, respectively. We find the following:
\begin{itemize}
\item[1.] Even including the higher-order contributions, the strong first-order EWPT still persists. However, $T_C$ and corresponding VEVs are altered into $T_C^{\text{Parwani}}=106.8$ GeV, $v_C^{\text{Parwani}}=201.5$ GeV, $v_{SC}^{\text{Parwani}}=1.2$ GeV, and $v_{SC}^{\text{sym,Parwani}}=144.8$ GeV in the Parwani scheme, while $T_C^{\text{CAE}}=107.8$ GeV, $v_C^{\text{CAE}}=202.7$ GeV, $v_{SC}^{\text{CAE}}=1.2$ GeV, and $v_{SC}^{\text{sym,CAE}}=145.3$ GeV in the CAE scheme. One can see that $v_C/T_C$ in both cases get lowered by about $10\%$. While the numerics may change in more refined analyses, such a tendency might be universal~\cite{Chiang:2018gsn}. 
\item[2.] $T$ dependences of $v(T)$ in the vicinity of $T_C$ in the Parwani and CAE schemes are milder than that in the HT potential case, which could reduce uncertainties in determining VEVs at $T_C$. Similarly, the behaviors of $v(T)$ and $v_S(T)$ above $T_C$ are also modified by the higher-order corrections substantially. We conclude that the HT scheme poorly describes the thermal history of VEVs. We note in passing that regardless of the calculation schemes, $v_S(T)$ asymptotically approaches to zero but does not vanish since $a_1\neq0$.
\item[3.] The scheme dependence of the thermal resummation is relatively small, which is in stark contrast to cases of EWPT in two Higgs doublet model~\cite{Cline:1996mga, Kainulainen:2019kyp}. In the current model, the tree-level potential structure, which is infrared safe, predominantly governs the first-order EWPT. On the other hand, thermal cubic terms derive it, but the thermal resummation screens contributions from those terms. { Incidentally, without thermal resummation, it is found that $T_C=104.9$ GeV, $ v_C=200.1$ GeV, $v_{SC}=1.3$ GeV, $v_{SC}^{\text{sym}}=143.3$ GeV in BP1, while $T_C=101.6$ GeV, $ v_C=206.6$ GeV, $v_{SC}=1.1$ GeV, $v_{SC}^{\text{sym}}=152.6$ GeV in BP2, which are not much different from those in Table~\ref{tab:EWPT}.}
\end{itemize}
Fig.~\ref{fig:EWPT126} show the results in BP2. Even though the numerics are somewhat different from BP1, the consequences found in BP1 all apply to this case as well. Therefore, we come to the conclusion that strong first-order EWPT in the degenerate-scalar scenario is possible in the both cases $m_{h_1}>m_{h_2}$ and $m_{h_1}<m_{h_2}$. Our findings are summarized in Table~\ref{tab:EWPT}.
%

%
%----------------------------------------------------------------------------------------------------------------------------------
\begin{table}[t]
\center
\begin{tabular}{|c|c|c|c|c|c|c|c|c|}
\hline
 & \multicolumn{4}{c|}{BP1} & \multicolumn{4}{c|}{BP2} \\ \hline
Scheme & HT & PRM & Parwani & CAE & HT & PRM & Parwani & CAE \\ \hline
$v_C/T_C$ & $\frac{184.4}{85.3}=2.2$ & $\frac{195.6}{78.2}=2.5$ & $\frac{201.5}{106.8}=1.9$ & $\frac{202.7}{107.8}=1.9$ & $\frac{191.3}{81.5}=2.3$ & $\frac{201.0}{74.8}=2.7$ & $\frac{206.3}{102.8}=2.0$ & $\frac{208.3}{104.4}=2.0$  \\ 
$v_{SC}$ [GeV]  & 1.5 & 1.2  & 1.2 & 1.2 & 1.3 & 1.1 & 1.1 & 1.1 \\ 
$v_{SC}^{\text{sym}}$  [GeV] & 134.6 & 137.3 & 144.8 & 145.3 & 144.2 & 146.6 & 153.2 & 154.1 \\ 
\hline
\end{tabular}
\caption{$T_C$ and the corresponding Higgs VEVs in BP1 and BP2 using the four calculation schemes. `HT' stands for the high-$T$ scheme in which the potential (\ref{VHT}) is used, `PRM' denotes the gauge-invariant calculation scheme proposed by Patel and Ramsey-Musolf~\cite{Patel:2011th}. Here we conduct the $\mathcal{O}(\hbar)$ calculation taking $\overline{\mu}=m_t$ as a reference value. `Parwani' refers to the ordinary one-loop calculation with the thermal resummation adopted in the work of Parwani~\cite{Parwani:1991gq}, while { `CAE' defines the same one-loop effective potential calculation but with the thermal resummation applied in the work of Carrington~\cite{Carrington:1991hz}  and  Arnold-Espinosa~\cite{Arnold:1992rz}}.}
\label{tab:EWPT}
\end{table}
%----------------------------------------------------------------------------------------------------------------------------------
Before closing, we comment on the case with $m_\chi=2$ TeV. In this case, one can find the first-order EWPT in the HT, Parwani, and CAE schemes while not in the PRM scheme. In the latter, $m_\chi$ has an upper bound arising from the condition (\ref{Tc_PRM}); that is, the right-hand side has to be lower than the left-hand side at zero temperature. Otherwise, the degeneracy point where $T_C$ is defined would not exist. However, in BP1, for instance, the right-hand side would exceed the left-hand side for $m_\chi\gtrsim 700$ GeV though the actual lower value is vulnerable to $\overline{\mu}$, which disallows one to define $T_C$. The DM mass bound in the PRM scheme is consistent with findings in Ref.~\cite{Chiang:2017nmu}.
It would be intriguing that to what extent this bound could be relaxed when one includes higher-order corrections omitted here.

%%%%%%%%%%%%%%%%%%%%%%%%%%%%%%%%%%%%%%%%%%%%%
% Conclusion
%%%%%%%%%%%%%%%%%%%%%%%%%%%%%%%%%%%%%%%%%%%%%
\section{Conclusion and discussions}\label{sec:con_dis}
We have investigated compatibility between the strong first-order EWPT and the degenerate-scalar scenario, which was advocated as a solution for evading DM direct detection experiments~\cite{Abe:2021nih}. For illustration, we considered the two benchmark points in which $m_{h_2}=124$ GeV and $m_{h_2}=126$ GeV taking the observed SM Higgs mass as $m_{h_1}=125$ GeV. We analytically showed that the suppression of $\sigma_{\text{SI}}$ driven by the smallness of $\delta_2$, which could be realized by a ratio of the mass difference of two scalars and the singlet VEV $v_S$, conflicts with one of the necessary conditions for the strong first-order EWPT. More concretely, in successful parameter space, sizable $\delta_2$ is mandatory, which requires that $v_S$ should be less than 1 GeV and the mixing angle $\alpha$ between the two scalars has to be nearly maximal, {\it i.e.,} $|\alpha|=\pi/4$ in order to compensate the suppression factor $m_{h_1}^2-m_{h_2}^2$ in $\delta_2$ [see Eq.~(\ref{del2})]. Moreover, the relative sign between $a_1$ and $v_S$ must be negative in order not to break the perturbativity bound of $d_2$ [see Eq.~(\ref{del2})]. Our numerical analysis based on \texttt{micrOMEGAs} also confirms that $\sigma_{\text{SI}}$ is not suppressed by the degenerated scalar masses. Nonetheless, the allowed regions are still present at around $m_\chi=62.5$ GeV and 2 TeV.
We also analyzed EWPT in the viable DM regions by exploiting four different calculation schemes: HT, PRM, Parwani, and CAE, where the former two methods are gauge invariant, while the latter two are ordinary gauge-dependent formalisms. All the calculations consistently indicate the strong first-order EWPT in the two representative benchmark points, BP1 and BP2.
In the case of $m_\chi=2$ TeV, four calculations except for the PRM scheme support the presence of the strong first-order EWPT. The PRM scheme cannot define $T_C$ properly in the heavy DM mass region due to the gauge-invariant vacuum energy of the electroweak vacuum higher than that of the symmetric vacuum at zero temperature.  
This result could change when the higher-order terms are taken into account, considering theoretical uncertainties in the $\mathcal{O}(\hbar)$ PRM calculation. 
Throughout the analysis, we have focused on $T_C$ and corresponding VEVs. As mentioned in Sec.~\ref{sec:ewpt}, however, nucleation temperature is more relevant for EWBG~\cite{Chiang:2017nmu,Baum:2020vfl,Biekotter:2021ysx}. In addition, CP violation is indispensable for baryogenesis. We defer them to future work.

% \begin{acknowledgments}
% \end{acknowledgments}

%%%%%%%%%%%%%%%%%%%%%%%%%%%%%%%%%%%%%%%%%%%%%%%%
%							References
%%%%%%%%%%%%%%%%%%%%%%%%%%%%%%%%%%%%%%%%%%%%%%%%

\bibliography{refs}

\end{document}